\documentclass[%
 reprint,
 amsmath,amssymb,
 aps,nofootinbib,   
]{revtex4-1}
\usepackage{bm}
\PassOptionsToPackage{linktocpage}{hyperref}
\usepackage[hyperindex,breaklinks]{hyperref}
\usepackage{enumitem}
\usepackage{slashed}

\renewcommand{\theta}{\vartheta}

\newcommand{\ket}[1]{\ensuremath{\left|\, #1\right>}}

\usepackage{array}
\usepackage{mathtools}

\usepackage{etoolbox}

\begin{document} 

\title{Saturon Dark Matter}

\author{Gia Dvali} 

\affiliation{
	Arnold Sommerfeld Center, 
	Ludwig Maximilians University, 	
	Theresienstra{\ss}e 37,
	 80333 Munich, Germany  
}
\affiliation{
	Max Planck Institute for Physics, 
	 F\"ohringer Ring 6, 80805 Munich, Germany
}


\begin{abstract}
  Saturons are macroscopic objects with maximal microstate 
 entropy.  Due to this property, they can be 
produced via quantum transitions from a homogeneous 
thermal bath, bypassing the standard exponential suppression characteristic of ordinary extended objects.  In this sense, saturons carry an advantage with respect to other macroscopic objects such as black holes and ordinary solitons.  Due to unsuppressed thermal production,  saturons can have interesting cosmological implications. In particular they can serve as viable dark matter candidates with some unique features.  Unlike ordinary particle dark matter,  
the superheavy saturons can freeze-in at very low temperatures.  
 A nucleation of a saturon can be described in terms of a saturated instanton.  This has implications for various phase transitions. 
 \end{abstract}


\maketitle

\subsection{Introduction} 

The origin of dark matter in the Universe is still a mystery. 
The especially valuable candidates are the ones with
calculable production mechanisms and observational signatures.
The known dark matter candidates, 
 can be divided in two general categories:  
  elementary particles and macroscopic composite objects. 
  Each category has certain universal characteristic properties. 
  
 The elementary particle dark matter is bounded in mass. 
  First, no quantum particle can be heavier that the Planck scale, $M_P$. 
  Secondly, the mass $m$ of a thermally-produced particle 
  at temperature $T$ is bounded from above by the temperature. 
  The production of a heavier particle 
   is exponentially suppressed by the Boltzmann factor, 
   $\exp(-m/T)$.
    
   The second category of dark matter includes  
  macroscopic objects. The examples are provided
  by black holes and various solitons. 
    Obviously, such dark matter ``particles"  can be much heavier than the Planck mass. The challenge however is the production mechanism. 
  
  In previous considerations, the macroscopic super-heavy dark matter is never produced via a direct quantum transition from a radiation thermal bath.       
    The reason is an exponential suppression
   characteristic of ordinary macroscopic objects.   
 We shall show that the origin
 of this suppression is entropic. To be more precise, it originates   
  from an insufficient entropy of ordinary extended objects
  as compared to the entropy of thermal radiation.   
  
Due to this,  the standard mechanisms for the production of 
macroscopic objects do not rely on their quantum materialization.   
For example, topological solitons can be produced during 
phase transitions via the Kibble
   mechanism \cite{Kibble:1976sj}.  Its variations for non-thermal phase-transitions can also be considered.    
   On the other hand, a creation of gravitationally bound 
   objects, such as  black holes, requires the presence of large inhomogeneities.  
        
  In the present article we shall propose a new class  
 of dark matter candidates, the so-called ``saturons" \cite{Dvali:2020wqi}. 
  In certain sense, these formations inherit the beneficiary features of both categories: 
 the classical composite objects and the quantum particles.      
 On one hand, a saturon represents a macroscopic object 
 of an arbitrarily large mass/size.  On the other hand, like an elementary quantum,  it can be produces thermally without an exponential suppression. 
 
The concept of saturons was introduced in \cite{Dvali:2020wqi}. 
However, several explicit examples have been constructed in 
  various field theories in \cite{Dvali:2019jjw, Dvali:2019ulr,
 Dvali:2021jto,  Dvali:2021rlf,  Dvali:2021tez, Dvali:2021ofp}. 
 The defining property of saturons  is that they saturate
 a certain upper bound on the entropy. 
 This bound is universally imposed by the unitarity. 
That is,  within the validity of a quantum field
theoretic description, the saturons are the objects of maximal microstate degeneracy.

   As shown in \cite{Dvali:2019ulr}, the phenomenon of saturation can take place  for objects with Lorentzian  as well as the Euclidean signatures.  In particular, it was demonstrated that  saturation 
   is fully applicable to instantons.  This has clear physical 
   consequences: upon a saturation, 
   the transition process mediated by the instanton saturates unitarity.   
  We shall refer to saturated instantons as ``{\it i}-saturons". 

  Due to the above special feature, saturons, 
 despite being the macroscopic entities, can 
 be produced via a quantum scattering from a thermal bath.  
  We shall explain that a formation of a saturon
  can be described as a quantum transition generated  by a saturated 
  instanton. That is, an {\it i}-saturon creates a saturon.   
   In fact, the saturated instantons can assist the
 phase transitions even if no stable Lorentzian saturon exists at 
 zero temperatures.      
  
 Due to their maximal entropy, the saturons share a number of striking similarities with black holes. 
  To be more precise, black holes 
 represent a special case of saturons.    
 However, unlike non-gravitational saturons, the black holes
 cannot be produced 
 via a quantum process from a homogeneous radiation thermal bath. 
   We shall explain this fundamental difference in detail.

    Due to the above feature, saturons can offer a qualitatively new 
 type of dark matter.  In particular, they can provide a superheavy macroscopic dark matter produced at unusually low temperatures via a quantum transition process.        
 
 In the present paper we shall discuss the essence of the above 
 phenomenon and outline various aspects of saturon cosmology. 
In particular, we discuss its implications for thermal transitions
and dark matter.

  \subsection{Saturons} 
  
   The term ``saturon" was introduced in 
 \cite{Dvali:2020wqi}  
    for describing an object exhibiting the maximal microstate degeneracy 
permitted by unitarity.  Various examples of such objects 
and their connection
with saturation of unitarity were given in series of papers
\cite{Dvali:2019jjw, Dvali:2019ulr,
 Dvali:2021jto,  Dvali:2021rlf,  Dvali:2021tez, Dvali:2021ofp}. 
For the detailed discussions, we refer the reader to 
the above articles.  Here, we shall briefly recount the 
key features of saturons that will be used in our analysis.
 First, we focus on saturons in space-times with a Lorentzian signature. 
 
  We consider an effective field theory  described by 
a set of degrees of freedom interacting via a dimensionless quantum coupling $\alpha$.  The validity of their 
field theoretic description implies 
that the coupling is weak,   $\alpha \ll 1$.
 
We assume that these degrees of freedom form a macroscopic composite object of mass/energy $E$ and 
a size $R$.  For example, the role of such object can be
played by a soliton or any other bound state.  
  
   Being composite, the object can be described by a mean occupation number of its constituents, which we denote by  
$n$. In the simplest cases the object has 
a single characteristic scale, $R$, which sets its size. 
 In such cases, the dominant contribution to the energy of the 
 bound state comes from the constituent quanta 
 of  wavelengths $\sim R$.  Each constituent contributes 
$\sim 1/R$ into the energy of the bound state.  
 Correspondingly, the total energy is, $E \sim n/R$. 
  
    For such objects, the  exact classical limit is well-defined and 
   reads, 
     \begin{equation} \label{Nlimit} 
  n \rightarrow \infty,~   n\alpha \, =\, {\rm fixed}, ~ R= {\rm fixed}\,.   
  \end{equation} 
 It is easy to see that the self-sustainability condition implies,
  $\alpha \sim 1/n$. In this case, the bound state can be 
  classically stable.   However, it can decay due to quantum effects 
  that are controlled by $1/n$ corrections.    
   Below, we shall work at finite but large $n$.

 As explained in \cite{Dvali:2020wqi}, an important universal characteristic of any bound state is the Goldstone scale.
  The reason is that macroscopic objects break spontaneously a set of symmetries, both space-time as well as internal. 
    
  Of course, the concept of spontaneous breaking of symmetry 
 becomes exact in the limit $n \rightarrow \infty$. 
  Only in this limit,  the Hilbert spaces built on 
  different degenerated vacua 
  are strictly orthogonal.      
   However, the description of objects in terms of spontaneous breaking of symmetry, represents an excellent approximation
 for large $n$. That is, the would-be Nambu-Goldstone modes 
 are gapless at least to leading order in $1/n$-corrections.  
 All the bound states at weak coupling $\alpha$, 
satisfy this condition.   
 
    Among various Goldstone modes, the Goldstone modes of 
    Poincare symmetry are especially useful. This is  due to their universal nature.  
  Obviously, any localized state breaks spontaneously the Poincare symmetry.   
 In particular, it breaks both space and time translations as well as the Lorentz boosts. Correspondingly, there exist a Goldstone boson 
 with a well-defined scale, which we denote by $f_G$. 
  It is easy to express this scale through $R$ and $n$, 
  as $f_G^2\, = \, n/R^2$. 
  
   In conclusion \cite{Dvali:2020wqi}, a generic bound state with a
 characteristic single scale $R$, admits an universal
 description in terms of parameters
$n$, $E$ and $f_G$.  They are uniquely expressed though 
$\alpha$ and $R$, 
 \begin{eqnarray} \label{N} 
  {\rm Number}: ~  n\, = \, \frac{1}{\alpha} , &&  \\
 \label{E}
 {\rm Energy} : ~  E \, = \, \frac{n}{R} \, =  \, \frac{1}{\alpha} \frac{1}{R}, &&  \\
  \label{fG}
 {\rm Goldstone~scale}: ~  f_G^2\, = \, \frac{n}{R^2} = \frac{E}{R} &&\,.
  \end{eqnarray} 

 It is important to understand that the above are the 
 universal relations for bound states formed by an interaction 
 with a scale $R$ and a weak coupling $\alpha$. 
 In particular, they are exhibited by all ordinary solitons.  
  A new feature that singles out a sub-class
   of saturons, is the maximal microstate entropy.  

 This feature has been widely considered 
 for black holes.   
  However, the microstate degeneracy is a property of 
  a much wider class of field theoretic objects. 
  This degeneracy can originate from various sources. 
  For example,  it can result from internal symmetries under which 
  the object transforms.

   Due to this, the composite states, such as solitons, 
   can be realised in a large number, $n_{st}$, of degenerate microstates.        
   The notion of a ``microstate" that we use, is standard. It amount to 
 distinct  quantum states that realize the same macroscopic 
  (classical)  properties of the object, such as the energy, size and a total angular momentum.  The states we refer to, are pure. 
  For such objects, a notion of the microstate entropy, defined as, 
  \begin{equation} \label{Sn}  
    S \, \equiv \, \ln(n_{st}) \,,  
 \end{equation}  
     is fully applicable. 
    
 An important question, investigated
 in papers \cite{Dvali:2019jjw, 
Dvali:2019ulr, Dvali:2020wqi} (for the summary, see \cite{Dvali:2021jto}),  is a maximal microstate degeneracy of a bound state.  
The answer is that an absolute upper bound on the microstate 
entropy of a bound state is given by the following 
three expressions  \cite{Dvali:2020wqi}, 
 \begin{eqnarray} \label{Smax1} 
  {\rm Area-form}: ~ S_{sat}  \, \sim  \, &&R^2f_G^2  \\
 \label{Smax2}
 {\rm Number-form}: ~ S_{sat} \,  \sim  \, && n  \\
   \label{Smax3}
 {\rm Coupling-form}: ~S_{sat} \,  \sim  \, && \frac{1}{\alpha}\,.
 \end{eqnarray} 
   The connection between these three formulations can be derived from the relations (\ref{N}), (\ref{E}), (\ref{fG}). 
  The saturons are defined as the objects exhibiting the above  maximal degeneracy.   
  
   Thus, from (\ref{E}) and  (\ref{Smax2}), (\ref{Smax3}),
  the saturon energy can be expressed through its entropy, 
  occupation number, or coupling,  as, 
  \begin{equation} \label{ES}
     E_{\rm sat}  \, = \, \frac{S_{\rm sat}}{R} \, = \,
    \frac{n}{R} \, = \, \frac{1}{\alpha} \frac{1}{R} \,. 
 \end{equation}  
           
   It has been established that attaining a higher degeneracy 
is not possible within the validity of any weakly-coupled set of degrees of freedom.
  That is, an attempt of obtaining a bound state with a microstate entropy exceeding $S_{\rm sat}$,  puts its constituent degrees of freedom out of the validity domain of the theory.     
 Several markers of this effect have been 
 identified in form of the breakdown of the loop-expansion
in coupling $\alpha$  \cite{Dvali:2019jjw, 
Dvali:2019ulr, Dvali:2020wqi}, as well as, in saturation of the unitarity bound 
by a set of scattering processes \cite{Dvali:2020wqi}.

   Notice  \cite{Dvali:2020wqi} that the first
relation in (\ref{ES}) coincides with the 
  well known Bekenstein bound on entropy \cite{BekBound}. 
  It emerges as a consequence of more stringent bounds
  (\ref{Smax1})  -  (\ref{Smax3}).

 Due to this reason, using this bound as a sole guideline for our purposes is not possible, as 
  it carries no information about the strength of the coupling
 (neither fundamental $\alpha$, nor collective  $n\alpha$)  or  about the occupation number of modes.  Correspondingly, it is not sensitive to the quantum field theoretic validity of the description and can formally be satisfied beyond its domain.   
 Therefore, we impose the necessary consistency condition that the system respects
the three bounds, (\ref{Smax1}), (\ref{Smax2}) and (\ref{Smax3}), simultaneously. This gives the relation (\ref{ES}) as a byproduct.  
 The saturon states saturate all three and thereby satisfy 
 (\ref{ES}).   
 
  The concept of saturation generlizes to objects with an Euclidean signature, such as instantons \cite{Dvali:2019ulr}. 
   For such entities all three bounds  (\ref{Smax1}) - (\ref{Smax3}) 
   are well-defined \footnote{In contrast, the formulation  
 of the Bekenstein bound for instantons is not possible, due to absence of the notion of energy.}. 
    For example, the instanton action exhibits 
   the degeneracy with respect to broken space and internal symmetries and the scale $f_G$ is unambiguously  defined. As explained in \cite{Dvali:2019ulr}, an useful intuitive way for picturing the instanton entropy is by using an auxiliary construct;  mapping an instanton  on a soliton in Lorentzian theory  with one additional  
 time-like dimension.  
 
   In general, upon saturation, the microstate entropy of an  instanton 
   becomes equal to is action.  That is, the instanton becomes an {\it i}-saturon.  This signals that the corresponding transition 
 process is close to saturating unitarity. 
 We shall come back to this point in the discussion of the role of saturons in thermal transitions.

 As shown in series of previous papers, 
 \cite{Dvali:2019jjw, Dvali:2019ulr, Dvali:2020wqi, Dvali:2021jto, Dvali:2021rlf,  Dvali:2021tez, Dvali:2021ofp},  
 saturons can come in variety of forms. 
 For example, they can be represented by topological or non-topological solitons, 
 baryons and other composite objects. 
 What unifies them is the saturation of the entropy bound. 
 
  Due to this, the saturons 
 share some striking similarities with black holes. 
 This is already apparent from the area-law form of their entropy
 given by (\ref{Smax1}).  This expression is identical to the Bekenstein-Hawking entropy \cite{BekE} 
  of a black hole 
 of radius $R$, 
 \begin{equation}\label{BHe} 
S_{\rm BH} \, \sim \,  R^2 M_P^2\,, 
\end{equation} 
 with the role of the scale of the Poincare Goldstone 
  $f_G$ played by the Planck mass,
 $M_P$. This is not surprising, since a black hole, just like any other 
 localized object, breaks the Poincare symmetry spontaneously. 
 The order parameter of this breaking is $M_P$.

As  explained by the microscopic theory of  
``black hole's quantum $N$-portrait" \cite{Dvali:2011aa}, 
in terms of the occupation number of gravitons and their coupling, 
the black hole entropy is given by the equations (\ref{Smax2}) and (\ref{Smax3}) respectively. 

It has been shown that saturons 
exhibit other striking similarities with black holes. 
These include a thermal like decay at initial stages as well as the 
scaling of the information-retrieval time
 in close similarity with the Page's time of a black hole \cite{Dvali:2020wqi, Dvali:2021jto, Dvali:2021rlf,  Dvali:2021tez, Dvali:2021ofp}.  
The relation between the maximal spin and the mass (or entropy) is 
also identical to black holes \cite{Dvali:2021ofp}.

  From the above perspective,  
  a black hole represents a saturon formed by a gravitational interaction. 
   As we shall see, surprisingly, this fact puts black holes at some disadvantage, as compared to saturons produced by other 
   interactions (e.g., the gauge forces). Namely, 
   a black hole cannot be produced from the homogeneous 
 thermal bath in the early Universe by means of a quantum transition. 
 In contrast, the non-gravitational saturons, can.

  \subsection{Saturon lifetime} 
  
 Before moving on, we comment on the saturon lifetime. 
First, saturons can be stable due to conserved charges, both topological and non-topological. 
Of course, in such cases the saturons with opposite charges must be
produced in equal numbers from the thermal bath. 

 However, even the saturons carrying no conserved 
 charges are macroscopically long-lived.   
 This is one of the many properties that saturons share with black holes.
 Similarly to a black hole, at the initial stages an unstable saturon decays via the emission of the soft quanta: 
 a quantum of energy $\sim 1/R$ is emitted per time 
 $R$.  Correspondingly, the timescale during which 
 an unstable saturon looses about half of its mass, is given by 
 $t_{\rm half} \sim S_{\rm sat} R$.

This timescale is very important as it represents a 
so-called ``quantum break-time" of the system. 
In particular, beyond this point the semi-classical approximation 
breaks down completely.
There are several effects contributing into this break-down. 
 
The first one is an ``inner entanglement", originally pointed out 
for black holes in \cite{Dvali:2013eja, Dvali:2021bsy}
and generalized to arbitrary high entropy states in 
\cite{Dvali:2018xpy, Dvali:2020wqi} (see also, \cite{Dvali:2022vzz}). 
This entaglement is developed due to inner rescattering 
of constituents and it reaches its maximum around 
$t_{\rm half}$.  We shall explain it in more details later 
on explicit examples of saturons.   
  
Secondly, 
There are indications \cite{Dvali:2020wft} that 
after $t_{\rm half}$,
 the decay slows 
down due to so-called ``memory burden" effect
\cite{Dvali:2018xpy}. 
The essence of the phenomenon is the following.

 In saturons, the degenerate microstates are
 achieved by excitations of gapless degrees of freedom, the so-called ``memory modes".   Their presence is universal in all saturons, while the emergence mechanisms can differ from case to case. 
 In a class of saturons (see examples below) the memory modes emerge as Goldstone modes of spontaneously broken symmetries
 \cite{Dvali:2019jjw, Dvali:2019ulr, Dvali:2020wqi, Dvali:2021jto, Dvali:2021rlf,  Dvali:2021tez, Dvali:2021ofp}. 
  
  In other cases they emerge as angular momentum modes that become gapless at the critical point \cite{Dvali:2017nis}. 
 As a result,  the saturon state is endowed with $\sim n $ distinct
 ``flavors" of gapless modes.  Each distinct microstate represents a 
 specific pattern of their excitations. Their number is exponentially large, resulting in the entropy $S \sim n$.  

  Due to the gaplessness of the memory modes,  the quantum information encoded in their excitation pattern, 
is energetically much cheaper than 
the analogous information carried by the asymptotic free quanta.  
Correspondingly, the saturon cannot decay without ``offloading" this information pattern.  Since there exist no gapless carriers 
among the free quanta, the offloading of information  is a highly suppressed process.  This effect tends to stabilize the saturon.

The full quantum evolution has only been studied 
numerically \cite{Dvali:2020wft}. These studies confirm 
the analytic estimates by showing that,  latest  by half decay, the process slows down.
   
   Due to the limitations of numerical analysis, the estimate of the 
  remaining  lifespan requires some guesswork.    
  On one hand, it is unlikely that a saturon with zero net charge 
 gets fully stable, as no conservation law forbids its decay. 
  On the other hand, it is evident that the memory burden 
 effect makes the lifetime much longer 
 than $t_{\rm half}$.

It is important to stress that the memory burden effect is an exclusive property of objects with enhanced capacity of information storage, 
such as saturons. This feature is not shared by composite objects of  low microstate entropy, such as ordinary solitons. 
 This sets a difference between  the decay processes 
 of these two classes of objects.  However, the difference 
 due to memory burden becomes significant after the 
mass of the object is reduced by order half.  

 The corpuscular resolution \cite{Dvali:2011aa} of classical objects, 
 such as black holes and solitons (for a corpuscular description 
 of  topological charges, see \cite{Dvali:2015jxa}) 
 as composites of high occupation number of quanta, 
 reveals close similarities in their quantum decays at initial stages.
 This decay can be understood as a result of
 quantum depletion due to the re-scatterings of the constituent quanta.  This mechanism is equally operative in black holes, 
 solitons and other field configurations \cite{Dvali:2011aa, Dvali:2017eba, Dvali:2017ruz, Dvali:2020wqi, Dvali:2022vzz, 
 Dvali:2021rlf,  Dvali:2021tez}.  
  In particular, in \cite{Olle:2020qqy} this picture has been applied to specific breathing scalar field configurations, called ``oscillons"  \cite{Copeland:1995fq}.  This study was motivated by 
 oscillons as dark matter \cite{Olle:2019kbo}.

    However, the similarities hold only up until the 
 half-decay time.  Beyond this point, exclusively for saturons,  the inner-entanglement and memory burden effects 
 set in.  No analogous effect is apparent for ordinary low entropy solitons \footnote{Of course, endowing the ordinary solitons, 
 such as, e.g., oscillons, with maximal microstate entropy, promotes them into the category of saturons, with all the consequences, such 
 as memory burden and inner entanglement.}.

   In summary,  the feature of a maximal entropy gives an advantage to saturons, as compared to ordinary solitons. 
   First, even in the absence of a conserved charge, saturons tend to get an extended lifespan via the memory burden effect.   
    This longevity shall be assumed in what follows. 
  Secondly, the maximal entropy allows for an unsuppressed quantum production of saturons from  a homogeneous thermal bath. 
  We now give a detailed discussion of this
  effect.

  \subsection{Thermal production of saturons}
  
   Saturons can certainly be produced by the conventional mechanisms characteristic of objects from the same universality class. For example, saturons of the type of topological defects 
 \cite{Dvali:2019jjw, Dvali:2019ulr, Dvali:2020wqi}, can be produced by the Kibble mechanism \cite{Kibble:1976sj}.  Analogously,  those saturons \cite{Dvali:2021tez, Dvali:2021ofp} that can be placed in the category of $Q$-balls \cite{Lee:1991ax, Coleman:1985ki},  can be produced by one of the mechanisms characteristic  for ordinary $Q$-balls \cite{Kusenko:1997si, Kusenko:1997vp, Dvali:1997qv}
(for a recent discussion, see, e.g., \cite{Bai:2022kxq}).

   In the case of production by such conventional mechanisms,  
 the novelty, as compared to ordinary solitons, 
  is that saturons can be produced in 
   exponentially large number of microstates and this affects  
   their subsequent evolution. In particular, it suppresses the
  rate of annihilation.  Correspondingly, a much higher number of saturons can survive as compared to ordinary solitons produced by the same mechanism.  This comparisons require a separate investigation which is 
  beyond the subject of the present paper. 
   
    Perhaps, the most interesting novelty, which we shall 
    focus on, is the possibility of
    a thermal production.  
   Ordinarily, the thermal production of extended objects, such as solitons,  is exponentially suppressed. A classic example illustrating 
  this suppression is the computation by Linde \cite{Linde:1981zj}, 
  who estimated the rate of  thermal production of the domain wall bubble at temperature $T$.   We start with the regime 
 of a ``thin-wall" approximation in which the bubble radius $R$ exceeds its thickness as well as the inverse temperature.  
  In this regime Linde's result amounts to,
  \begin{equation} \label{Linde}
    \Gamma(T) \sim T^4 \left (\frac{A_3}{T}\right )^{\frac{3}{2}} \, 
    {\rm e}^{-\frac{A_3}{T}} \,,
   \end{equation} 
  where $A_3$ is a three-dimensional Euclidean action  which
  essentially is equal to the energy of the bubble, 
  $E_{\rm bub}$ \footnote{We changed notation of 
 \cite{Linde:1981zj} in order
  to use different symbols for the action ($A$) and the
entropy ($S$).}.    
    
  For a saturated  bubble, the thin-wall approximation is 
  applicable only order of magnitude wise. 
 We shall therefore estimate the rate by a more general method, 
 which does not rely on such approximation.  As a consistency check,
 we show that our estimate agrees 
  with Linde's result. 
   
 Since the goal of our analysis is to point out  a qualitatively new 
 mechanism, we shall be limited by an order of magnitude estimates. The unimportant numerical factors will be set equal to one. 
    Of course, in applications to realistic models,  more precise estimates must be performed. 

   We assume that a theory admits a solitonic object 
  of some characteristic size $R$ and energy 
  $E_{\rm bub} $. For simplicity, we shall assume 
  a spherical symmetry, although the arguments extent beyond this 
  approximation.  The role of the object in particular can be played 
  by a bubble separating the two vacua, as in
 Linde's analysis. However, our estimate applies to
 a larger class of objects.    
  
   Let us consider a thermal bath at temperature $T$.  For the beginning, we limit ourselves with a single type of scalar species in thermal equilibrium.   Then, the energy density of the thermal bath is 
$\sim T^4$, whereas the entropy density is given by $\sim T^3$.    
   Correspondingly,  the energy ($E_{\rm rad}$) and the entropy
 ($S_{\rm rad}$) of a spherical region of radius $R$ are given by,
  \begin{equation} \label{STandET}
    S_{\rm rad}(T) = (RT)^3\,, ~ \, E_{\rm rad}(T) =  T^4 R^3 \,.   \end{equation} 
        
 Let  us estimate the probability for quantum transition of this spherical region into a state of the soliton (e.g., a vacuum bubble) 
 of the same radius $R$. 
     
  We shall assume that the soliton is created out of a thermal region 
  of the equal radius.  Although different regimes are also 
  possible, as we shall explain, they do not change the estimate substantially. 
 Then, in order for the transition to take place, the soliton energy must 
not exceed the energy of the thermal bath within the same
sphere: 
  \begin{equation} \label{ElessT}
   E_{\rm bub}  \leqslant  E_{\rm rad}(T)\,.   
   \end{equation} 
 Taking into account (\ref{STandET}), the 
 ``on-threshold" production of a bubble gives a condition,
   \begin{equation} \label{condition}
   \frac{E_{\rm bub}(T)}{T} =   S_{\rm rad}(T)\,.   
   \end{equation}   
 Let us first assume that the bubble carries a zero (or a very small) entropy. This is an implicit assumption in conventional 
discussions.  
 In particular, this is the case in Linde's analysis. 

   Then, creation of such a bubble represents 
 a transition from a high entropy state
   (radiation) into a low-entropy one (soliton). 
  The corresponding rate is suppressed by the entropy factor, 
      \begin{equation} \label{expS}
    \Gamma(T) \, \sim \, 
    {\rm e}^{-S_{\rm rad}(T)} \,.
   \end{equation} 
 Taking into account (\ref{condition}), it is clear that 
 the expression  (\ref{expS}) recovers the
 exponential factor given in (\ref{Linde}).  
 The difference is that we did not use any thin-wall approximation.  
  
  The above result explains in clear physical terms why 
  the thermal production of ordinary solitons is exponentially suppressed.  It is highly improbable for the thermally-distributed 
 particles to get organized into a coherent state of a very low entropy, such as the soliton.

   The same picture explains why the situation with 
  saturons is fundamentally different. 
   The reason is that the saturon entropy is maximal
  for a given quantum field theoretic system.  
  In particular, it can exceed  
 the entropy of the radiation bubble. 
  Correspondingly, the suppression factor can be compensated
  by the entropy of the saturon, 
       \begin{equation} \label{expS}
    \Gamma(T) \, \sim \, 
    {\rm e}^{-S_{\rm rad}(T) + S_{\rm sat}(T)}\,.
   \end{equation}     
  Of course, the formation probability cannot exceed one. 
  In the extreme case, the saturon formation will saturate the unitarity 
  of the time-evolution.    
  
  Let us discuss this phenomenon in more details. 
 For generality, we consider a thermal bath with 
  $N_*$ particle species in thermal equilibrium.  Then, the entropy and 
  the energy of a radiation bubble of radius $R$ are given by,   
   \begin{equation} \label{STandETN}
    S_{\rm rad}(T) = N_*(RT)^3\,, ~ \, E_{\rm rad}(T) =  N_*T^4 R^3 \,.   \end{equation} 
    In order for a saturon to form, the following two conditions must be satisfied, 
     \begin{equation} \label{SEcondN}
    S_{\rm rad}(T) \leqslant S_{\rm sat}\,, ~~~ \,
   E_{\rm sat} \leqslant E_{\rm rad}(T)\,.
        \end{equation} 
   Taking into account (\ref{STandETN}) 
 and (\ref{E}), we obtain the following 
 window of temperatures for the formation of a saturon, 
     \begin{equation} \label{TRsat1}
      \left(\frac{S_{\rm sat}}{N_*} \right )^{\frac{1}{3}}  \gtrsim
      TR
        \gtrsim  \left(\frac{S_{\rm sat}}{N_*} \right )^{\frac{1}{4}} \,.
        \end{equation} 
Expressing the saturon entropy through the occupation number 
of its constituents, $n$, via the relation (\ref{Smax2}), we can rewrite the above expression as,   
      \begin{equation} \label{TRsat2}
      \left(\frac{n}{N_*} \right )^{\frac{1}{3}}   \gtrsim TR
        \gtrsim  \left(\frac{n}{N_*} \right )^{\frac{1}{4}}\,.
        \end{equation}    
 Notice that the window exists only 
 for 
 $\frac{S_{\rm sat}}{N_*} \, \gtrsim \, 1$ (equivalently, 
 for $\frac{n}{N_*}  \gtrsim 1$), which shall be assumed.  
    
   In fact, as we shall explain later, 
  the saturon formation probability is peaked at the lower bound of this inequality.  That is, the transition probability 
  is maximized when the saturon mass exactly matches
   the energy of the radiation within the same sphere. 
 In other words,  the optimal transition happens for,   
   \begin{equation} \label{Tformation}
 E_{\rm sat} \, = \, E_{\rm rad}(T) \, = \, N_*T_s^4R^3 \,
 \end{equation}
  Taking this into account, 
 we obtain that the optimal temperature $T$ for the formation of 
  the saturon of radius $R$ is,  
     \begin{equation} \label{RandTopt}
     T \sim \frac{1}{R}  \left(\frac{S_{\rm sat}}{N_*} \right )^{\frac{1}{4}}
        \sim  \frac{1}{R} \left(\frac{n}{N_*} \right )^{\frac{1}{4}}\,.
        \end{equation}   
     Correspondingly, using the expressions (\ref{E})  and (\ref{ES}), the saturon  energy can be expressed as the function of the optimal temperature and the saturon entropy (or the occupation number), 
          \begin{equation} \label{EandTopt}
     E_{\rm sat}  \, \sim \, T  \, S_{\rm sat}^{\frac{3}{4}}N_*^{\frac{1}{4}}
        \sim  \,T  \,  n^{\frac{3}{4}}N_*^{\frac{1}{4}} .
        \end{equation}  
                      
  We must note that while at the optimal temperature saturons
 can be produced out of thermal bath without an exponential suppression, the process cannot equilibrate easily.
  First, if the saturon entropy 
  is higher than the entropy of radiation, the inverse process
 is exponentially suppressed. 
   The second factor is the expansion rate of the Universe 
 which suppresses the probability of equilibration.

  \subsection{Abundance of saturon dark matter}
  
  \subsubsection{Bottom-up: estimate of required suppression} 
  
  Saturons may compose a part or an entirety of the dark matter 
  in the Universe.  In the latter case, we can put 
  a bottom-up constraint on the probability of their production. 
   
   Let $p$ denote a total probability of the transition of a thermal 
   sphere of radius $R$, into a saturon that occupies the same 
   sphere.   The transition rate is maximal around the optimal temperature and quickly falls off beyond it.   
   Thus,  below $T$, we shall end up on average with one saturon 
  per volume $V \, = \, R^3/p$. 
  Taking into account the condition (\ref{Tformation}), 
   we get the following
   energy density of saturons at optimal formation temperature  $T$,
  \begin{equation} 
  \rho(T)\, = \,   
  p N_*T^4 \,.
 \end{equation}
  After this moment, the saturon energy density red-shifts as 
  matter, and for todays temperature $T_0 \simeq 2.4 $K, reaches the value, 
   \begin{equation} 
  \rho_{\rm sat}(T_0) \,  = \, pN_*TT_0^3 \,.
 \end{equation}
    Equating this to the current dark matter density,
    $\rho_{\rm DM}$, we get the 
    following  condition on $p$,  
  \begin{equation} \label{p1}
  p \, = \, \frac{\rho_{\rm DM}}{N_*TT_0^3} \,.
 \end{equation}
 In terms of the present day radiation energy 
  density, $\rho_{\rm rad}(T_0) \, \sim  \, T^4_0$, we write,  
  \begin{equation} \label{p2}
  p \, \sim \, \frac{\rho_{\rm DM}}{\rho_{\rm rad}(T_0)} 
 \frac{T_0}{T} \frac{1}{N_*} \,
 \sim \,  \frac{\rm eV}{N_* T} \,.
 \end{equation}
  For example, saturons produced around the 
  temperatures $T\sim$ MeV, can serve as dark matter candidates provided 
 they satisfy $p\sim 10^{-7}$.

 \subsubsection{Top-down: probability of production}  
 
  Let us now estimate the probability of production, $p$, in 
  top-down approach. 
  We consider a thermal bath at temperature $T$ in an expanding
  Universe.  We are interested in the probability of production 
  of a saturon of radius $R$ from the spherical domain of the same 
  radius filled with radiation. 
   The saturon formation process can be viewed as the 
   transition of thermal quanta of occupation number 
   $n_{\rm rad} = N_*(TR)^3 = S_{\rm rad}$ into a coherent state of saturon 
   of occupation number $n \, = \, S_{\rm sat}$.  
   Since $S_{\rm rad} \leqslant S_{\rm sat}$, we have $n_{\rm rad} \leqslant n $. 
   
   That is,  the transition process 
   represents a conversion of $n_{\rm rad}$ thermally distributed particles into $n$ coherently organized
   quanta of softness $1/R$.  
 Mentally, this can be viewed as a scattering process
   \begin{equation} \label{process} 
    n_{\rm rad} \rightarrow n \,, 
  \end{equation} 
    obtained by summing over different transition amplitudes. 
The momentum transfer per particle is $\sim 1/R$. 
Correspondingly, the characteristic time-scale of the process is $t_{tr} \sim R$.  We shall call it a ``transition-time".

  The transition takes place for the optimal temperature (\ref{Tformation}), 
 for which the energy of radiation 
 $E_{\rm rad}(T)$  matches the saturon mass, 
 $E_{\rm sat}$. 
Let us denote this optimal temperature by $T_s$. 

    For  $T \neq T_s$, the saturon production is suppressed.     
   Indeed, for  $T < T_s$, one needs to produce 
  a saturon  out of a thermal sphere 
   larger than its size. This causes an extra exponential suppression. 
    For, $T > T_s$, the system 
   produces a state that is not a saturon ground state and correspondingly carries a small fraction of the available entropy. 
   
    This physical intuition is supported by the analysis of the scattering amplitudes \cite{Dvali:2020wqi},  which shows that 
 the probability for the saturon production in a scattering 
 process at center of mass energy $E$, quickly diminishes 
 outside of the following window, 
  \begin{equation} \label{Edelta}
  \Delta E \,  \equiv \,  |E -  E_{\rm sat}| \lesssim 
  \frac{E_{\rm sat}}{S_{\rm sat}} \, .
 \end{equation}
 In the present case, $\Delta E$ must be taken as the
allowed spread of the thermal energy within the
sphere of radius $R$ that is transiting into a saturon state.

 Now, due to the expansion of the Universe, the temperature 
  drops and the energy of radiation within the radius $R$ gets red-shifted. 
  Correspondingly,  the window of opportunity during 
 which $\Delta E$ satisfies  (\ref{Edelta}), is open only for 
a finite time $\Delta t_s$.   
   If this time is shorter than the transition time $t_{tr} \sim R$, the probability  gets suppressed.
     
    The time $\Delta t_s$ can be easily estimated. 
    First,  the energy within the sphere
   is related with the change of temperature, $ T \, = \, T_s + \Delta T$,
   as,  
  \begin{equation} 
  \frac{\Delta  E}{E_{\rm sat}} \sim \frac{\Delta T}{T_s} \,.
  \end{equation}  
   The time interval required for this change is, 
  \begin{equation} 
   \Delta t_s  \sim \frac{\Delta T}{T_s H} \, .
  \end{equation}  
  Using (\ref{Edelta}),  we get the optimal time interval,  
  \begin{equation} 
   \Delta t_s  \sim \frac{1}{S_{\rm sat}  H} \, .
  \end{equation} 
 For convenience, we define a parameter that controls the suppression, 
    \begin{equation} 
  \beta \equiv  \frac{\Delta t_s}{t_{tr}} \sim 
    \frac{1}{S_{\rm sat}HR} \, .
  \end{equation} 
When $\beta \gtrsim 1$, in absence of additional suppression factors,  
the transition probability is close to one, 
$p \sim 1$. In the opposite case, $\beta \ll 1$, the probability 
is suppressed. To leading order, we can estimate  $p \sim \beta$.  

  Using the relation (\ref{RandTopt}) and the formula 
  for the Hubble (\ref{Hubble}), we can express the parameter
  $\beta$ as, 
      \begin{equation}  \label{betaS}
  \beta \, \sim  \,  \frac{M_P}{T_s}\frac{1}{(N_* S_{\rm sat}^5)^{\frac{1}{4}}} \,. 
  \end{equation} 
  Finally, remembering the relation (\ref{Smax3}) between the entropy
  of the saturon  and the coupling, we can write, 
        \begin{equation}  \label{betaA}
  \beta \, \sim  \,  \frac{M_P}{T_s}\frac{\alpha^{\frac{5}{4}}}{N_*^{\frac{1}{4}}} \,. 
  \end{equation} 
  
   We wish to notice the following curios thing.
  In scenarios with $\beta \ll 1$, we can fix the 
  required entropy by equating the expressions (\ref{p2}) and 
  (\ref{betaS}). 
  Then, the temperature drops out of the equation and the 
  proper entropy is fixed as,   
    \begin{equation} \label{Spbeta}
  S_{\rm sat}  \, \sim \,   \left( \frac{M_P}{\rm eV} \right )^{\frac{4}{5}}
  N_*^{\frac{3}{5}} \sim 10^{22}  N_*^{\frac{3}{5}}\,.
 \end{equation}
 
  We shall now apply the above knowledge to 
 cosmological production of saturons. 
 We distinguish two cases, to which we 
 refer to as ``freeze-out" and ``freeze-in"
 scenarios.  The names are borrowed from the terminology  
 of ordinary particle dark matter. In particular, the term 
 ``freeze-in"  was introduced in \cite{Hall:2009bx}.
  
  However, these names require some clarifications.
 Unlike the case with a particle dark matter, saturons are never in thermal equilibrium.  Rather, the terms ``freeze-out" and ``freeze-in"
refer to the state of the constituent quanta at the time of the saturon 
production.  
  In the ``freeze-out" scenario the constituents are in thermal 
equilibrium, whereas in case of  ``freeze-in",  they are not. 
  This is the reason for the name choices.   
  However, the physics is fundamentally different.
  This is due to the fact that saturons are collective phenomena, 
and their production mechanisms are not reducible to
the ones of ordinary particle dark matter.

     \subsubsection{Freeze-out} 
   
    In a freeze-out scenario, the saturon constituent species are in thermal equilibrium at the moment of the saturon production. This puts certain conditions on the parameters. 
    
      First, the temperature must satisfy $T \gtrsim 1/R$, where 
      $R$ is the size of a saturon.  
   Since the energy of each constituent is $\sim 1/R$, for a lower temperature their numbers would be Boltzmann-suppressed.     
       
    Secondly, the re-scattering rate of the saturon constituents, 
          \begin{equation} \label{2to2}
    \Gamma_{2\to 2} \, \sim \, \alpha^2 T \,,
   \end{equation} 
    where $\alpha$ is the coupling,  must be larger  than the Hubble expansion rate,
          \begin{equation} \label{Hubble}
    H \, \sim \, N_*^{1/2} \,
    \frac{T^2}{M_P} \,.
   \end{equation} 
 This condition puts the following constraint on the coupling, 
         \begin{equation} \label{2to2OUT}
    \alpha \, \gtrsim \,  N_*^{1/4} \,
    \left (\frac{T}{M_P} \right )^{1/2} \,.
   \end{equation}       
   Using (\ref{Smax3}), this equation can be rewritten as the bound 
   on the saturon entropy:
       \begin{equation} \label{SFr}
      S_{\rm sat} \, \lesssim  \, \frac{1}{N_*^{\frac{1}{4}}}\left(\frac{M_P}{T} \right)^{\frac{1}{2}}\,.
   \end{equation}  
    Correspondingly, through (\ref{E}) and (\ref{RandTopt}), the above translates into
the following  bound on the saturon mass, 
        \begin{equation} \label{EFr}
      E_{\rm sat} \,  \lesssim\, N_*^{\frac{1}{16}}T^{\frac{5}{8}}M_P^{\frac{3}{8}} \,.
   \end{equation}  
   
 Taking into account the relation (\ref{2to2OUT}), 
 in the expressions (\ref{betaS}) and (\ref{betaA}),
  we get the following lower bound on the parameter $\beta$, 
         \begin{equation} 
  \beta \, \gtrsim \,  \left(\frac{M_P}{T} \right)^{\frac{3}{8}} 
  N_*^{\frac{1}{16}} \,. 
\end{equation} 
Thus, in  a freeze-out scenario, the probability of saturon formation 
at the optimal temperature can be  $p \sim 1$. 

 This finding should not be understood as the rejection 
 of the freeze-out scenario for a saturon dark matter. 
  There exist additional factors that can easily reduce the saturon density to an acceptable level.
  
    First,
the saturon entropy can be slightly below the unitarity upper bound. 
  Since the sensitivity of $p$ with respect to the saturon entropy 
  is exponential, this situation is rather natural.  
   
 Secondly, the saturon density can be reduced by the 
 additional stages of reheating.  For example, due to
 a relatively late decay of heavy particle  species. 
  Such a decay injects an additional energy into radiation and correspondingly dilutes the density of the saturon matter.

     \subsubsection{Freeze-in}

     In a freeze-in scenario, the saturon constituents are not in thermal equilibrium.  At the time of saturon production, their abundance 
     is negligible.  
   This requires that their re-scattering rate (\ref{2to2}) 
   is lower  
     than the Hubble rate (\ref{Hubble}). 
   Correspondingly,  we obtain,   
         \begin{equation} \label{2to2IN}
    \alpha  \, \lesssim \,   N_*^{1/4} \,
    \left (\frac{T}{M_P} \right )^{1/2} \,.
   \end{equation}

 Using (\ref{Smax3}), this imposes the following lower bound 
   on the saturon entropy, 
         \begin{equation} \label{SFrIN}
      S_{\rm sat}  \,  \gtrsim \, \frac{1}{N_*^{\frac{1}{4}}}\left(\frac{M_P}{T} \right)^{\frac{1}{2}}\,.
   \end{equation}     
   As discussed, the optimal temperature for saturon production is (\ref{RandTopt}).  
   We can estimate $\beta$ under the assumption that the   
number freezes
 around this epoch.

 Taking into account the inequality (\ref{2to2IN}) 
 in (\ref{betaA}),  we get the following lower bound on the parameter $\beta$ in freeze-in scenario, 
         \begin{equation} 
  \beta \,  \lesssim \,  \left(\frac{M_P}{T} \right)^{\frac{3}{8}} 
  N_*^{\frac{1}{16}} \,. 
\end{equation} 
 Since the right hand side of the equation is larger than one, 
 this bound is less informative than in the freeze-out scenario. 
 Thus, in freeze-in scenario, the probability 
 suppression has to be calculated on a case by case basis.     
    
    In a freeze-in scenario, the mass of a generic saturon 
    is limited essentially only by the energy of the entire
 Hubble patch.  For example, solar mass saturons can 
 freeze-in below the QCD temperatures.

    \subsection{$SO(N)$ examples}
    
    For purely illustrative purposes, 
 we shall give two examples of saturon states
   realized in theories with large symmetries such as $SO(N)$  (or 
    $SU(N)$). They shall emerge as the bound states of solitonic and non-solitonic nature respectively. Many variations of the models are possible.

  \subsubsection{Vacuum bubble saturon}

    As the first example, we construct a saturon as a vacuum bubble in
  $SO(N)$ theory.  For this, we slightly modify the 
     model  of \cite{Dvali:2020wqi, Dvali:2021tez}, using the same 
     basic mechanism for achieving the entropy saturation of the bubble.  
 The key idea \cite{Dvali:2019jjw, Dvali:2019ulr, Dvali:2020wqi} is to have 
 solitons with spontaneously broken large global symmetry in their  interior.   The general conclusions are not limited to 
a particular model.
  In \cite{Dvali:2020wqi, Dvali:2021tez} these solitons were chosen 
 in form of the vacuum bubbles, interpolating between the unbroken and broken phases. 
     
  We shall construct a similar model with a slightly simplified 
  representation content.
  The model consists of two real scalar fields: $\phi_j$,
 transforming as $N$-dimensional representation under 
 $SO(N)$ symmetry  (with $j=1,2, ...,N$, the $SO(N)$ index)
  and a singlet $\chi$. The Lagrangian
 has the following form,    
\begin{eqnarray}
    \label{SO(N)Model}  
 L \, =\,  \frac{1}{2} \left(\partial_{\mu} \phi_j \right)^2\, + \,  \frac{1}{2} \left(\partial_{\mu} \chi \right)^2 \, - && \nonumber  \\
  - \, \frac{\alpha}{4} \left (f_G^2
  - \phi_j^{2} - \chi^2 \right)^{2} - \frac{\alpha}{2} \phi_j^{2}\chi^2 \,.                  
\end{eqnarray}
 where $\alpha$ is a coupling and $f_G$ is a scale. 
 In addition to  $SO(N)$, the theory also possesses a discrete $z_2$ symmetry under which
 $\chi \rightarrow -\chi$.  This symmetry is not essential for our 
 considerations and can be explicitly broken by 
additional terms. 
 
 Notice that unitarity forces the following bound 
 on the parameters of the theory \cite{Dvali:2019jjw, Dvali:2019ulr, Dvali:2020wqi}, 
 \begin{equation} \label{ANbound}
 \alpha N \lesssim 1\,.
 \end{equation}
  The product $(N \alpha)$ plays the role analogous to 
  't Hooft coupling \cite{planar}.   
 For values exceeding the bound (\ref{ANbound}), the  
 $\phi_j$ a $\chi$ fields are no longer the valid degrees of freedom. 
   In particular, this is signalled by the breakdown of the 
  loop expansion in power series 
  in $(N \alpha)$. Any discussion of objects obtained 
  as trustable solutions in terms of $\phi_j$ and $\chi$ fields,
  is bounded by (\ref{ANbound}).  This fact plays a fundamental role 
in limiting the entropy of such objects  \cite{Dvali:2019jjw, Dvali:2019ulr, Dvali:2020wqi}. 
 
   The theory has the two sets of degenerate vacua. One is the vacuum 
   with unbroken $SO(N)$-symmetry, 
   $\phi_j =0, \chi= f_G$. This vacuum is doubly degenerate 
   with respect to the $z_2$ symmetry.  In this vacuum the theory
 exhibits a mass gap of order $m \equiv \sqrt{\alpha} f_G $ for all the fields. 
   
   In addition,  there exists a vacuum $\phi_i^2 = f_G^2, \chi=0$, in which 
   $SO(N)$ is spontaneously broken down to $SO(N-1)$. 
   Correspondingly, the spectrum includes  $N-1$ massless Goldstone bosons with the decay constants 
   given by $f_G$.  In addition, the vacuum houses 
  two fields with masses $m$: the $\chi$-boson
   and a radial mode of the $\phi_j$-field. 
   
     Other extrema of the potential are:  the
  maximum at $\phi_j = \chi =0$  and the saddle points
  at $\phi_i^2 = \chi^2 = f_G^2/3$ which connect the vacua
  with $SO(N)$ and $SO(N-1)\times z_2$ symmetries.  
  
      Following  \cite{Dvali:2020wqi, Dvali:2021tez}, the saturon solution can be obtained from the bubble of an $SO(N-1)$-invariant 
      vacuum embedded in $SO(N)$-invariant one.     
     In order to understand the structure, we first consider a
a spherical bubble in a thin-wall regime. This is the regime when the radius of the bubble, $R$,  is much larger than its thickness $\sim 1/m$. 
We remark preemptively that 
for the saturated bubbles one has $R \sim 1/m$. Correspondingly, the  
thin-wall approximation only works as an 
order of magnitude estimate.  Nevertheless,  extrapolation from this regime is useful for understanding the saturon properties. 

    For a thin-wall bubble the absolute 
    values are represented by the  functions 
    \begin{equation}
    \label{profiles} 
  \sqrt{\phi_i^2} = \rho(r,t), ~  |\chi| = \rho_{\chi}(r,t) \,,
\end{equation}
where $r$ is a radial coordinate. 
  The function $\rho(r)$ interpolates from $\rho( 0 )\simeq f_G$ to $\rho(\infty) = 0$ over an interval $R$, which 
determines the size of the bubble.
In fact, for the thin-wall bubble, 
$R \gg 1/m$, the transition mostly happens within the bubble-wall 
which has a thickness $\sim 1/m$.

The function $\rho_{\chi}(r,t)$  has the opposite asymptotic behaviour, 
$\rho_{\chi}(0)\simeq  0,~ \rho_{\chi}(\infty) = f_G$,  interpolating  from almost a zero value at the bubble center to one
of the $z_2$-degenerate vacua at $r=\infty$. 
Also here,  the transition takes place within the bubble wall. 
          
   Depending whether it carries an overall $SO(N)$ charge, the 
 bubble can be either stationary or oscillatory. In both cases the lifetime is macroscopic in $N$. 
   In both cases, the bubble exhibits a large degeneracy, due to the spontaneous breaking of the $SO(N)$-symmetry in its interior.
  Obviously,  due to $SO(N)$-symmetry of the theory,  
  any bubble obtained by the $SO(N)$-transformation 
   of the solution, is also a solution.  This  results in a microstate
   degeneracy. 
 The  bubbles that saturate the entropy bound exist in the regime 
   $\alpha N \sim 1$, which according to 
   (\ref{ANbound})  is also an unitarity bound on the 
   description in terms of degrees of freedom 
  $\phi_j$ and $\chi$.     

From the analysis of \cite{Dvali:2020wqi, Dvali:2021tez}, it is easy 
to see that for a saturon bubble, which is the main focus of our interest, $ m \sim 1/R$. The occupation number $n$  
of its constituents is 
\begin{equation} \label{nNA}
n \simeq  N \simeq \frac{1}{\alpha} \,. 
 \end{equation} 
 The microstate degeneracy of a bubble
 is given by a binomial coefficient, 
  \begin{equation} \label{nstN}
 n_{st} \simeq \frac{(n+N)!}{n!}{N!} \sim  \left(1+\frac{N}{n}\right )^N 
\left (1+ \frac{n}{N} \right)^n  \,, 
 \end{equation} 
 where we have used Stirling approximation for large $n$ and
 $N$.  At the saturation point (\ref{nNA}) the entropy is 
      \begin{equation} \label{SsatBN}
        S_{\rm sat} \simeq N \simeq \frac{1}{\alpha} \,.
     \end{equation}              
  The saturated bubble represents a bound state of $n \sim N$ quanta of wavelengths $R \sim m^{-1}$.  It is clear that the bubble satisfies all the properties (\ref{N}),  (\ref{E}), (\ref{fG}), as well as the expressions for the saturon entropy,  (\ref{Smax1}),  (\ref{Smax2}), (\ref{Smax3}).

    As shown in \cite{Dvali:2021tez},  there exist bubbles for which 
   the functions $\rho(r), \rho_{\chi}(r)$ are time-independent. 
     
   In the present model, for a stationary bubble, the field $\phi$ has the following 
   form, 
    \begin{equation}
    \label{eq:rotation}
    \phi \, = \, \hat{O}^{\dagger} \Phi \, ,
\end{equation}
where 
\begin{equation}
    \label{eq:phi} 
    \Phi \, = \, \rho(r) \big[1, 0, \dotsc, 0 \big] \,,
\end{equation}
and
\begin{equation}
\hat{O} \, = \, \exp\mspace{-3mu} \big[
                i\,\omega t \,\hat{T} \big] \,.
\end{equation}
  Here  $\omega$ is the frequency and $\hat{T}$ is one of the broken generators of $SO(N)$. 
Thus, the bubble spins, but only in an internal $SO(N)$-space.  
  
  The functions $\rho(r)$ and $\rho_{\chi}(r)$ again
 have the opposite asymptotics,   
  $\rho(0)\sim f_G \,~\, \rho(\infty)\, =\, 0$ and  
 $\rho_{\chi}(0) \, < \, f_G\, ~ \, \rho_{\chi}(\infty) \, = \, f_G$. 
 Notice that $\rho(0) \, = \, f_G$ and $\rho_{\chi}(0) = 0$ are realized 
 only for $R = \infty$ limit.

 The classical stability of the solution is due to the fact that 
 it is energetically favourable to store the $SO(N)$ charge
 in form of a bubble rather than in form of free particles. 
In this sense, the bubble is similar to a $Q$-ball
\cite{Lee:1991ax, Coleman:1985ki}.  However,  it carries
a maximal microstate entropy which makes 
all the difference in the present context.

From the analysis of \cite{Dvali:2020wqi, Dvali:2021tez}, it is easy 
to see that for an internally-spinning saturon bubble
(\ref{eq:rotation}), 
$\omega \sim m \sim 1/R$ and the occupation number $n$  
of its constituents satisfies (\ref{nNA}). 
Correspondingly, its entropy is given by (\ref{SsatBN}).

   \subsubsection{Example of scalar force bound states} 
   
   In our second example, the spontaneous breaking 
   of global $SO(N)$-symmetry is not required. 
   Instead, the bound state is formed due to an attractive
  force mediated by a scalar $\pi$. The Lagrangian is, 
  \begin{eqnarray}  
      \label{PIModel}  
 L = \frac{1}{2} \left(\partial_{\mu} \phi_j \right)^2\, + \,  \frac{1}{2} \left(\partial_{\mu} \pi \right)^2 - \frac{1}{2} m_{\pi}^2\pi^2 \, - && \nonumber  \\
 - \frac{1}{2} m^2 \phi_j^{2}  \,   
 - \, \sqrt{\alpha}\, m\, \pi \phi_j^{2} \, + \, ... \,,                  
\end{eqnarray}
where again $\alpha$ is the coupling and $m, m_{\pi}$ are masses. 
 We assume that the energy is bounded from below due to 
 the higher order non-linear terms, not shown explicitly. 
As in the previous example, the validity domain of the description 
is bounded by (\ref{ANbound}).   

 The scalar $\pi$ mediates an attractive Yukawa type force with strength $\alpha$. The attractive potential energy 
between two non-relativistic $\phi_j$ particles is 
\begin{equation}
V_{\pi}(r) = - \alpha \frac{{\rm e}^{-m_{\pi} r}}{r}\,.   
\end{equation}  
  For $m_{\pi} \lesssim m$, this force 
can produce a bound state of many non-relativistic 
  $\phi_j$ quanta. The  size of the bound state is given by 
 $R \sim 1/m$, whereas the occupation number of quanta in the bound state is 
 \begin{equation} \label{nalpha}
n \simeq 1/\alpha \,.
\end{equation} 
 This can be seen by 
 balancing, at distance $r \sim 1/m$,
 the kinetic energy of each quantum of de Broglie 
 wavelength $\simeq 1/m$, with the attractive 
 potential energy from the rest, 
   \begin{eqnarray}  
      \label{kinpot}  
 (E_{kin} \simeq m)  \,  \sim \, (E_{pot} \simeq (n \alpha) m)  \,.                  
\end{eqnarray} 
 This gives (\ref{nalpha}). 
   The mass of the bound state is, 
      \begin{eqnarray}  
      \label{massPI}  
 E_{\rm sat}  \sim m n \, \sim  \frac{m}{\alpha}  \,.                  
\end{eqnarray} 
 Notice that such a bound state exists even for a single 
 $j$-flavor of $\phi_j$, i.e.,  for  $N=1$.  

However, the existence of many species creates 
an exponentially large microstate degeneracy. 
Due to $SO(N)$-symmetry, the bound state has the same mass 
for arbitrary set of flavors.  For fixed $n$ and $\alpha$, the degeneracy increases exponentially 
with $N$. For example, a symmetrized state of $n$ quanta, 
has a degeneracy given by (\ref{nstN}). 

 As in the previous example, due to unitarity bound (\ref{ANbound})  and 
 relation between the occupation number and coupling
 (\ref{nalpha}),  the maximal entropy of the bound state  is reached for 
(\ref{nNA}) and is given by (\ref{SsatBN}). 
  For this choice of parameters, the bound state becomes a saturon.
  In a sense, the bound state is similar to the atomic nuclei 
 in a theory of large number of nucleon flavors.   
 
  As in the previous example, the degeneracy can be 
  understood in terms of Nambu-Goldstone phenomenon 
  of spontaneously broken $SO(N)$ symmetry by the saturon 
 bound state.     
 
  As already discussed, in general, the saturon bound states are macroscopically long lived. 
 They can also be exactly stable. 
  For the bound states transforming non-trivially under $SO(N)$, 
  the stability can be understood by the conservation of 
 $SO(N)$-charge, which costs less energy in the bound state
 as compared to the state of free quanta.   
   
 However, even $SO(N)$ invariant bound states,  
 can be very long-lasting.  
    The long timescale of quantum decay is a 
 general feature of the saturated bound states of $N$ quanta.
 The first reason is a highly suppressed rate of re-scattering
(see, \cite{Dvali:2020wqi, Dvali:2021tez}).   
  In addition, the decay is slowed down by the 
  memory burden effect \cite{Dvali:2018xpy, Dvali:2020wft}.  
    As a result, even saturons carrying no conserved charges 
can be sufficiently long-lived for being dark matter. 
 We shall comment more on this later.

\subsection{General bounds}    

We wish to study the above examples
of $SO(N)$-saturons in cosmological context. 
    However, since they include a large number $N$
  of particle species, we must take into account some 
  general constraints.

\subsubsection{Species bound} 

    First, we must take into account that the
  gravitational cutoff is lowered to a so-called ``species" scale 
  \cite{Dvali:2007hz},
   \begin{equation} \label{sp}  
   M_{\rm sp} \equiv \frac{M_P}{\sqrt{N}} \,.   
\end{equation}  
 This constraint is imposed by black hole physics 
 and thereby is fully non-perturbative in nature.
 In particular, it cannot be avoided by any sort of a 
 fine-tuning of radiative corrections 
 or by a resummation of perturbative series. 
   Notice that due to low cutoff, the theories with large $N$ are motivated by the Hierarchy Problem \cite{Dvali:2007hz}.    
  
   For the present purposes, we must keep in mind that 
   the species scale $M_{\rm sp}$ limits from above the temperature 
   of the Universe.  
   
  It is important to notice that although the scale  (\ref{sp}) represents an upper 
  bound on the masses of elementary particles, it does not 
  restrict the masses of composite objects such as saturons or black holes. 
    Correspondingly,  even the superheavy saturons can be produced at very low temperatures.   

 This is one of the features that makes the present case very different from the particle dark matter scenarios with many species 
 such as discussed, e.g., in  \cite{Dvali:2009fw}  or  in \cite{Cohen:2018cnq}.

   \subsubsection{Bounds from BBN}
   
 Other important bounds on the parameters of  the $SO(N)$ 
 sector come from the big bang nucleosynthesis (BBN). 
   
   First, the abundance of new species during BBN must be 
   negligible.  
Secondly, the BBN epoch must be within a so-called ``normalcy"
temperature.  This concept was introduced in
\cite{Arkani-Hamed:1998sfv}  in the contexts of cosmological 
production of Kaluza-Klein graviton species in the theory of large extra dimensions \cite{Arkani-Hamed:1998jmv}. 

 The essence 
of this constraint is that in the BBN epoch, 
the thermal bath 
must 
cool-down  predominantly due to the expansion of the Universe. 
In the presence of the additional light species, 
a new cooling channel opens up. The bath can cool-down 
via re-scatterings of the ordinary particles into the new species. 
 In order not to affect the successful predictions of 
 BBN, the rate of such a process must be less than the
 Hubble expansion rate.  
  
 If we denote the coupling between the Standard Model 
 and $SO(N)$ species by  $\alpha_N$, the normalcy constraint on 
 the temperature reads, 
 \begin{equation} \label{normalcy1}
   T \alpha_N^2N \,  \lesssim\, \sqrt{N_*} \frac{T^2}{M_P} \,. 
 \end{equation}
 Notice that unitarity puts a constraint similar to (\ref{ANbound}) on the coupling,  
 $\alpha_N N < 1$.   For the coupling at its upper bound, 
 the normalcy temperature must satisfy, 
  \begin{equation} \label{normalcy2}
   T \, \gtrsim  \, \frac{M_P}{N\sqrt{N_*}}\,. 
 \end{equation}
  Taking $T$ of order the BBN temperature, $\sim$ MeV,  we get the 
  following bound on $N$ for   
 species with $m < T$,  
   \begin{equation} \label{BBNnormN} 
   N \,  \gtrsim \, 10^{22} \,.
 \end{equation} 
   Of course, for  $m \gg $ MeV, there is no BBN constraint on 
   $N$, since the production of species
 is exponentially suppressed.  
 
 \subsubsection{Star cooling bounds} 
 
 Another potential constraint on the light particle species 
comes from the processes of star cooling.
Of course, this constraint only applies if 
the masses of new species are lower than the 
temperature in a star. 
  For a star of core temperature $T$, the cooling rate 
due to emission of species is  
 given by the left hand side of (\ref{normalcy1}). 
 In each particular case we must make sure that this 
 rate is much less than the rate of ordinary cooling induced by the emission  of the Standard Model quanta such as the neutrino.

\subsection{Thermal production of $SO(N)$-saturons}

We now estimate the rate of a thermal production
 of $SO(N)$ saturons.  
  We start with a freeze-out scenario.

    \subsubsection{Freeze-out} 
    
  We consider a thermal bath of particles formed over the
 vacuum with unbroken $SO(N)$-symmetry.
 This applies to both examples of saturons. 
 We assume that the temperature is 
 $T \sim m$, and that the particles are still in 
 approximate thermal equilibrium. 
In theory (\ref{SO(N)Model}) the vacuum bubbles can form 
by the Kibble mechanism, provided the Universe start out at 
temperatures $T\gg m$ above the symmetry restoration point.  
   We are not interested in such a scenario. 
 Instead, we shall study the saturon formation directly 
 from the thermal bath as a quantum 
 transition process.   
     
  Since we are working at large $N$, for definiteness, let us assume that the total number of species is dominated 
  by the saturon constituents,
   i.e., 
   \begin{equation} \label{NstarN}
   N_* \simeq N \,.
 \end{equation}           
     Then,  the entropy and the energy of the thermal bubble of radius $R$  are given by (\ref{STandET}) with $N_*=N$,              
    \begin{equation} \label{STandETO(N)}
    S_{\rm rad}(T) = N(RT)^3\,,~~~ \, E_{\rm rad}(T) =  NT^4 R^3 \,.   \end{equation} 
    Since we also have $T \sim 1/R \sim m$, the condition (\ref{condition}) of the equality between  the energies of the saturon bubble and the corresponding thermal sphere, automatically implies the equality of their entropies,  
        \begin{equation} \label{SsatSradO(N)}
    S_{\rm sat} =  S_{\rm rad} = N \,.
       \end{equation}     
 Taking into account  (\ref{SsatSradO(N)}) and 
 the general formula (\ref{SFr}), we get that 
  $N$
 and $T$ must satisfy, 
          \begin{equation} \label{SFrN}
       N \,   \lesssim \, \left(\frac{M_P}{T}\right)^{\frac{2}{5}}\,.
   \end{equation}  
   Correspondingly, we get the following bound on the 
   saturon energy
      \begin{equation} \label{EFrN}
      ~ E_{\rm sat} \,  \lesssim \, T^{\frac{3}{5}} M_P^{\frac{2}{5}}\,.
       \end{equation}  
  The thermal production of saturons  stops below 
   the temperature $T\sim 1/R \sim m$ and the saturon number quickly freezes out.

    \subsubsection{Freeze-in}  
     
     We now discuss a freeze-in scenario.
     In this case we assume that saturon constituents are not in 
     thermal equilibrium. 
     Thereby, their re-scattering rate must be lower  
     than the Hubble rate.  According to (\ref{SFrIN}), this 
      implies, 
           \begin{equation} \label{SFrNin}
     N  \, \gtrsim \, \frac{1}{N_*^{\frac{1}{4}}}\left(\frac{M_P}{T} \right)^{\frac{1}{2}}\,.
   \end{equation}     
   Expressing via (\ref{SsatSradO(N)}) 
  the saturon entropy through $N$, from general formula 
  (\ref{EandTopt}) we get the following expression for saturon 
  energy, 
       \begin{equation} \label{EFrNin}
     E_{\rm sat} \, \sim  \,  T N^{\frac{3}{4}}N_*^{\frac{1}{4}}\,.
       \end{equation}  
 The bound (\ref{SFrNin}) translates as the lower bound on this energy, 
     \begin{equation} \label{boundEFrNin}
     E_{\rm sat} \, \gtrsim \, N_*^{\frac{1}{16}} 
     T^{\frac{5}{8}} M_P^{\frac{3}{8}}\,.
       \end{equation}

    Assuming that to the leading order the 
 probability suppression is $p \sim \beta$,
 and applying the equation (\ref{Spbeta}) to the present case, 
  we fix $N$ as,          
       \begin{equation} \label{NfromBBN}
  N  \, \sim \, 10^{22}  N_*^{\frac{3}{5}}\,.
 \end{equation}
 
 \subsubsection{Examples} 
 
 {\bf Example 1} \\
 
  Let us take the optimal temperature $T \sim $ TeV and assume 
  that only the Standard Model species are in equilibrium.  This implies $N_* \sim 100$.
   Then, from (\ref{NfromBBN}) we get, $N \sim 10^{23}$, 
  the saturon size, $R \sim 10^{-12}$cm and energy 
   $E_{\rm sat} \sim 10^{20}$ GeV.
   
  Notice that the population of free species in $SO(N)$ sector is
  negligible.  Their production rate remains subdominant to 
  the Hubble rate throughout the cosmological history.    
  The $SO(N)$-sector contributes into the energy density 
  of the Universe in form of the 
  saturon bound states that constitute dark matter.     
    \\

 {\bf Example 2} \\

  Let us find out a phenomenologically acceptable maximal value 
  of $N$. This is achieved by maximizing the number of 
  species $N_*$ that are in thermal equilibrium at the freeze-in temperature.  Of course, these species  cannot 
  belong to $SO(N)$ sector.  They
 can remain in thermal equilibrium latest by BBN and must decouple 
 before this epoch.  We shall therefore assume 
 that their masses are $m_* \sim $MeV.   
  Due to the unitarity constraint analogous to 
  (\ref{ANbound}),  their coupling is bounded from above  
 by $\alpha_* \sim 1/N_*$.
 Since, by assumption, these species are in thermal equilibrium, their scattering 
 rate must exceed the Hubble rate prior to BBN, which puts a constraint, 
 \begin{equation} \label{NstarBBN}
   N_* \, \lesssim \, (M_P/T_{BBN})^{2/5} \sim 10^{9} \,.
\end{equation} 
   Taking $N_* \sim 10^9$, 
   from (\ref{NfromBBN}) we obtain $N \sim 10^{28}$. 
    The corresponding mass of the saturon (\ref{EFrNin}) is, 
       \begin{equation} \label{ENBBN}
     E_{\rm sat} \, \sim \,   10^{20} {\rm GeV} \,.
       \end{equation}     
   Notice that this saturon is a macroscopic object of the size 
       \begin{equation} \label{ENBBN}
     R \,  \sim \, 10^{-6} {\rm cm}  \,,
       \end{equation} 
   which exceeds its own Compton wavelength by $28$ orders of magnitude.  This is due to the fact that it represents a bound state 
 of $N \sim 10^{28}$ particles.  
 
  This scenario barely satisfies the star-cooling constraint. 
 However, the question about  extra $N_*$ species 
 must be addressed. They drop out of thermal equilibrium 
 before BBN. If stable, they will contribute into dark matter density
 together with saturons.

       Notice that the above scenario is fully consistent with the
       species bound (\ref{sp}).   For $N \sim 10^{28}$, the gravitational 
       cutoff is about $M_{\rm sp} \sim 100$TeV.  
   Correspondingly, the freeze-in temperature
  somewhere above or around BBN, is safely below it.  As already discussed 
  \cite{Dvali:2007hz}, due to a low cutoff, the theories with many species have an independent motivation from the point of view of the 
 Hierarchy Problem.  In this class of theories, the possibility of saturon dark matter comes as an extra bonus.

 \subsection{Comment on inner entanglement} 
 
 As we have already discussed, in both  $SO(N)$ examples the stability of certain saturons 
  is due to $SO(N)$ charge. A saturon, which represents a bound state of  $\phi$-quanta of occupation number
 $n \sim N$,  
   forms a  representation of the $SO(N)$ group. Due to the binding potential, the saturon mass is less  than the energy of free $\phi$-quanta 
  that match the $SO(N)$ quantum numbers of the saturon.    
  Basically, a saturon is stabilized by the $SO(N)$-charge in a way similar to a stabilization of ordinary structures,
  such as nuclei or planets, by a conserved baryon number. 
The new feature is that the microstate degeneracy of the 
object saturates the bounds (\ref{Smax1}) - (\ref{Smax3}).     
  
  Due to this property, even a system of saturons with opposite quantum numbers is macroscopically long lived. That is, a
 system consisting of a saturon and an anti-saturon placed on top of each other, decays very slowly. This is due to $1/N$-suppression
 of the annihilation process.  
 
 The energy of a saturon-anti-saturon pair 
is by $\Delta E \sim nm$ lower than the energy of 
$2n$ free quanta. Correspondingly, the decay products 
must contain by $\sim n$ less free $\phi$-particles
than the initial number of constituents, which is equal to $2n$.  
  That is, during the decay process, the number of quanta must be reduced by $\sim n$.  
 
  However, an explosive annihilation of large number of constituents 
  into a small number of quanta, 
 a so-called ``quantumization" process \cite{Dvali:2022vzz},  is exponentially suppressed. 
 Notice that the suppression of quantum transition matrix elements  
 between the states of large and small numbers of particles can be derived from 
 a very general argument \cite{Dvali:2020wqi}.  
 Correspondingly, the dominant process is a gradual decay due to annihilation/scattering  of small number of constituent quanta into 
the asymptotic ones
\cite{Dvali:2011aa, Dvali:2013eja, Dvali:2017eba, Dvali:2017ruz, Dvali:2021rlf,  Dvali:2021tez, Dvali:2022vzz}.

 During the decay process, the saturons are subject to generation of an inner 
 entanglement. Originally this effect was pointed out 
 within the quantum $N$-portrait resolution of black holes
 and of de Sitter state as composites of $N$ gravitons \cite{Dvali:2013eja, Dvali:2017eba}. 
  However, the inner entanglement takes place in generic 
  objects with enhanced capacity of information storage 
  (enhanced microstate entropy) and it goes hand in hand 
  with the memory burden effect 
  \cite{Dvali:2018xpy, Dvali:2020wft, Dvali:2021bsy, Dvali:2022vzz}. 
    
   In the present case, this is due to the internal rescattering 
   of $\phi$-quanta, without leaving the bound state.  
  For example, a pair of quanta of a flavor $j$ scatters into
   a two-particle state  entangled with respect to the flavor index, 
\begin{equation} \label{EntL} 
    \ket{\phi_j, \phi_j'}  \, \rightarrow \, 
   \frac{1}{N}\sum_l \ket{\phi_l, \phi_l'} \,,
 \end{equation} 
   where prime indicates the distinction with respect to other quantum numbers such as the angular momentum.  
   This process contributes into the generation of an inner entanglement 
 which reaches its maximum  approximately after time  $\sim N R$.  
  Beyond this point, the system represents a maximally entangled 
  $n$-particle state. Not even approximately, such a state can be treated semi-classically.

   \subsection{Comparison with black holes}

   It is illuminating to compare the non-gravitational saturons to  black holes.  The key difference is that a quantum production 
   of a black hole from a homogeneous thermal bath is 
   strongly suppressed.  Instead,  for a cosmological black hole production,  
  it is necessary that some other mechanism prepares  
  large density inhomogeneities which then collapse into black holes
  \cite{Escriva:2022duf}. 
   
   This difference may sound somewhat surprising, since a black hole 
  represents a particular case of a saturon. The specifics is 
  that it is a bound state produces by the gravitational interaction. 
  Interestingly, this property makes a black hole  
 production from a radiation thermal bath
 highly improbable. 
 
 The ``secret" is that, unlike other saturons, 
in case of a black hole one and the same interaction (gravity)  governs the two important time-scales: \\

{\it 1)} a would-be quantum transition-time, $t_{tr}$; 

and  

{\it 2)}  the Hubble expansion time of the Universe. \\

    In order to see this,  we  apply the previous reasoning to a quantum creation  of a black hole of radius $R$, from the 
    thermal sphere of the same radius,
 in a homogeneous radiation dominated Universe.     
  We first assume that $R$ is much smaller than the 
  Hubble radius $R_H$. In this regime, the effects of the 
  Hubble curvature can be ignored. 
  
     For such a black hole, the mass, $M_{\rm BH}$, and the radius, 
  $R$, satisfy the standard flat-space relation: 
    \begin{equation} \label{RBH}
   R \,  \sim  \, \frac{M_{\rm BH}}{M_P^2} \,.   
   \end{equation}     
   We equate this mass to the initial energy of radiation within
 the equal-size sphere, 
      \begin{equation} \label{MBHRAD}
   M_{\rm BH} \, \sim \, N_* T^4R^3\,.   
   \end{equation} 
 Next, dividing both sides of the equation by $M_P^2$ and taking into account the relation  (\ref{RBH}), 
 we get, 
       \begin{equation} \label{MBHRAD}
  R^2 \,  \sim   \, \frac{M_P^2}{N_* T^4}\, \sim  \, R_{H}^2\,.   
   \end{equation} 
  The scale  $R_H$ is the Hubble radius.

   Thus, the minimal radius of a thermal sphere that matches the energy of the same size black hole, is the Hubble radius.
   The ratio of black hole to radiation entropy is, 
   \begin{equation}\label{SbhSrad}
    \frac{S_{\rm BH}}{S_{\rm rad}} \, \sim \, \frac{M_P}{\sqrt{N_*}T} \,.
   \end{equation} 
    Since, the temperature is always less than the species 
    scale (\ref{sp}), this number is larger than one. 
   Thus, the entropy of a Hubble size  black hole is always higher than the entropy of radiation.

  However, for a Hubble size  black hole the 
  flat space analysis breaks down, since the Hubble expansion must be taken 
  into account. 
    This expansion prevents the black hole formation. This
  is due to the following two factors.  
    
    First,  a would-be quantum transition time, 
   $t_{tr} \sim R$, is equal to the Hubble time. 
  Thereby, during this time the expansion of the Universe dilutes the energy 
within the initial Hubble sphere.

 Secondly, the Hubble patch has the Gibbons-Hawking entropy
\cite{Gibbons:1977mu},
\begin{equation}\label{GH} 
S_{\rm GH} \, \sim \,  (R_H M_P)^2\,.
\end{equation} 
 Notice that this entropy is equal to the Bekenstein-Hawking entropy 
 (\ref{BHe}) of a flat space black hole of radius $R=R_H$.  However, we are not 
 in a flat space and a would-be black hole is substituted by 
 the Hubble sphere. 
 That is, in an expanding Universe, the  most entropic state of the size of the Hubble horizon is the Hubble patch itself.
 
  The system, therefore, has no incentive for creating any other state, 
  since the saturon in form of the Hubble patch is already present.
  In other words, the dominant fraction of the Hilbert space is 
  taken up by the microstates of the Hubble patch.  
    As a result, no black hole formation takes place.

  Let us now consider an alternative possibility. 
  Since black holes come in all possible sizes, 
 one can think of a quantum process 
 of spontaneous collapse of a radiation 
 sphere 
 into a sub-horizon black hole. 
 Such processes are exponentially suppressed.  
 In order to see this, let us consider a quantum collapse of a spherical region of radius $R \ll R_H$ into a black hole of the same mass.  For $R \ll R_H$ the radius of an equal mass black hole satisfies  $R_{BH} \ll R$.  Its Bekenstein-Hawking 
   entropy  is, 
   \begin{equation} \label{Bekenstein}
    S_{\rm BH} \, \sim \, N_*T^4R^3 R_{BH}  = 
    S_{\rm rad} \,  (R_{BH}T) \,.
\end{equation} 
 For $R_{BH} \gg T^{-1}$, this exceeds the initial entropy of the radiation sphere.
    Nevertheless, this entropy is not sufficient for compensating 
    the exponential suppression of the
  quantum collapse due to the size difference.    
            
    The Euclidean action over a most symmetric trajectory 
    leading to such a collapse can be estimated as $S_E \sim N_*(TR)^4$.  
   This exceeds the black hole entropy by a factor 
    $\sim R/R_{BH}$.  Correspondingly, the excess of the black hole 
    entropy over the entropy of radiation is insufficient for 
    compensating  
  the exponential suppression of  the quantum collapse. 
   The over-all suppression factor,  
  \begin{equation}
     {\rm e}^{-S_E  +  S_{\rm BH} - S_{\rm rad}} \, = \, 
      {\rm e}^{- (\sim  S_{\rm BH} R/R_{BH})} \,,
  \end{equation}    
 renders the process highly improbable.

    Thus, in a homogeneous Universe a radiation thermal bath
  cannot collapse into a black hole spontaneously.      
     This goes in contrast with generic saturons.
  As we have seen, they can materialize via quantum transitions  
     from a radiation thermal bath.
   This difference is due to the following two factors: 
   1) the universal nature of the gravitational interaction, and
   2)  the fact that 
    a non-gravitational binding force can be much stronger than gravity.  Correspondingly,  a saturon that is formed by 
  such a force, can be much lighter than 
    a same-size black hole.  Due to this, it can easily 
    match the size and the energy of the 
    thermal bath within a sub-horizon sphere.   
     In this respect, saturons are similar to other 
  extended objects, such as solitons.  
    
   At the same time, the saturons share with black holes the property 
   of maximal entropy, which the ordinary objects do not posses.  Therefore, while matching the energy 
  and the size of a thermal sphere, a saturon dominates the entropy count. 
   
    In other words, saturons inherit those features 
   from both classes of objects that allow for unsuppressed quantum formation in  
   a radiation dominated universe.

  \subsection{Saturated instantons and enhanced bubble nucleation} 
  
      So far, we were discussing the saturons defined as objects 
      in space-times with Lorentzian signature, either at zero or 
      at finite temperatures.    
     However, as shown in \cite{Dvali:2019ulr}, this concept 
   fully generalizes to Euclidean field configurations, such 
   as instantons.  Of course,  for such entities 
   the notion of energy is not defined.  However, the bounds 
  (\ref{Smax1}) - (\ref{Smax2}) do not rely on energy. 
  Correspondingly, these bounds are well defined for instantons. 
 
 The microstate entropy 
 of an instanton is defined according to the 
  degeneracy of its Euclidean action. 
   As shown \cite{Dvali:2019ulr}, the saturation 
 of the bounds  (\ref{Smax1}) - (\ref{Smax2}) by this entropy, has 
 a trasparent physical meaning.     
    
   In particular,  the existence of a saturated  instanton, 
   an {\it i}-saturon,   
  signals  the saturation of unitarity by the corresponding transition
  process. 
    A violation of the bounds  (\ref{Smax1}) - (\ref{Smax2}) by an instanton is not possible as this would invalidate the
respective quantum field theoretic description.       
     For example, it has been demonstrated that 
  in large-$N$ gauge theories, the instantons saturate the entropy bound when the theory approaches the regime of confinement \cite{Dvali:2019ulr}.    
  
     A saturated instanton describes a creation of a saturated 
  object.   In short: \\

  {\it a transition generated by an {\it i}-saturon creates a saturon.}  \\
  
   This effect can have interesting cosmological implications,
  regardless dark matter.  
    In particular,  it can affects the dynamics  of the high temperature phase transitions. 
    For example, the rate of the bubble nucleation at finite $T$
  can be dramatically enhanced by Euclidean {\it i}-saturons,  even if no stable Lorentzian saturon 
   exists in the theory at $T=0$.  
  Instead, it suffices that a critical bubble at finite $T$ is a saturated state. 
    The existence of such a bubble goes hand in hand with the existence of a saturated instanton that creates it.   
    
     We illustrate this on explicit example of 
    a thermal nucleation of a saturated bubble.     
   We consider a saturated vacuum bubble introduced in 
        \cite{Dvali:2020wqi}. 
      The model consists of a scalar field  $\hat{\phi}$ transforming 
      in adjoint representation 
   of $SU(N)$-group.  The Lagrangian density is given by,
    \begin{equation} \label{LagN}
  L = \frac{1}{2} {\rm Tr} (\partial_{\mu} \hat{\phi} \partial^{\mu} \hat{\phi}) -
    V(\hat{\phi}) \,, 
  \end{equation}
 where the zero temperature scalar potential has the following form, 
    \begin{equation} \label{adjoint}
    V(\phi) \, = \, \frac{\alpha}{2} {\rm Tr} \left ( f\hat{\phi}  - (\hat{\phi}^2 - \frac{\hat{I}}{N}{\rm Tr}\hat{\phi}^2 )\right )^2 \,.
  \end{equation}
Here, $\hat{I}$ is the unit $N\times N$ matrix and 
$\hat{\phi(x)}$ is an $N\times N$ traceless hermitian matrix. 
The parameter $\alpha$ is the coupling and $f$ is the scale. 
As in the previous large-$N$ examples, the validity of the description 
imposes the bound (\ref{ANbound}) on the 't Hooft coupling
$\alpha N$.  
 
 At zero temperature, the theory has a set of degenerate vacua. 
 They consists of a $SU(N)$-symmetric vacuum, as well as the   
 vacua in which the $SU(N)$ symmetry is spontaneously
 broken down to various maximal subgroups,  $SU(N-K)\times SU(K)\times U(1)$ with  $0< K < N$. 
 
  At non-zero temperature, the vacua are no longer 
degenerate. In particular at $T \rightarrow \infty$,  only  the
$\hat{\phi}=0$ vacuum is present. 
 We shall assume that the degeneracy is lifted 
 by the additional terms in a way that,
 below some critical temperature, the broken symmetry vacuum 
is lower than the unbroken one. In such a case the system can tunnel 
via a bubble nucleation. 

 For definiteness, we assume that the order parameter 
 of the transition is oriented along the following component,  
  \begin{equation} \label{TVEV}
  \hat{\phi} \,   = \,  \phi(x) \, {\rm diag} ((N-1), -1, ....,-1) 
  \frac{1}{\sqrt{N(N-1)}} \,.    
  \end{equation}
   That is, the transition takes place from 
   $SU(N)$-symmetric vacuum to 
   $SU(N-1)\times U(1)$-symmetric one.

Then, the temperature dependent potential of the order parameter 
$\phi$ can be arranged  
in the following form, 
     \begin{equation} \label{adjointT}
    V(T, \phi) \, = \, \frac{\alpha(T)}{2} \phi^2\left ( f(T)  - \phi \right )^2 \,  \, - \, \mu^2(T) \phi^2 \,.  
   \end{equation} 
Of course, the effective parameters 
$\alpha(T),  f(T)$ and  $\mu(T)$ depend on $T$. 

 The parameter $\mu^2 > 0$ introduces a bias, which 
 lowers the free energy of the broken symmetry vacuum 
 $\phi \neq 0$ below the symmetric one $\phi =0$.   
The thin-wall approximation corresponds to  $\mu^2  \ll 
m^2 $,  where  
  $m^2\, \equiv \,  \alpha \,  f^2$ is the effective mass of the $\phi$ 
  excitation at temperature $T$.  

   The saturated bubble and the corresponding saturated 
instanton exists for  $\mu  \sim m$.  
Correspondingly, the thin-wall 
approximation is valid order of magnitude wise. 
 It therefore makes sense to use it as a starting point 
 and approach the parameter regime of saturation from this side. 
 
    The effective potential (\ref{adjointT}) has two minima, 
    \begin{equation} \label{VEVsT}
\phi = 0\,  ~  {\rm and}~ \, \phi  \, \simeq \,  f \left (1 + \,  \frac{2 \mu^2}{m^2} \right ) \,.
\end{equation} 
 The energy split between them is 
    \begin{equation} \label{SplitE}
  \epsilon \simeq f^2\mu^2 \,.
\end{equation} 

   The transition between the vacua takes place via a bubble 
   nucleation. The nucleation rate is exponentially suppressed 
   by the Euclidean action 
   of the instanton $A_{inst}$ that materializes a critical  bubble, 
       \begin{equation} \label{Ainst}
  \Gamma(T) \sim  {\rm e}^{- A_{inst}}\,.
\end{equation}  
 We now estimate this action. 
 We first find the radius $R$ of the critical bubble, 
 via extremizing  $A_{inst}$.   
  
  For $T \gg R^{-1}$, 
   the rate of the bubble nucleation is given by 
  the equation  (\ref{Linde}). 
   In this regime, the action effectively factorizes \cite{Linde:1981zj},
        \begin{equation} \label{A43}
  A_{inst} \, \rightarrow \,  \frac{A_3}{T} \,, 
\end{equation}   
   where the three-dimensional part, 
  $A_3$,  coincides with the energy of the static bubble.    
   
    In the thin-wall approximation the energy of a static bubble of radius 
   $r$ is given by, 
     \begin{equation} \label{A3}
  A_3 = E_{bub} = -\frac{4}{3}\pi r^3 \epsilon \, + \, 4\pi r^2 A_1\,,
\end{equation} 
  where  $A_1$ is the tension of the bubble wall. 
   For a thin-wall bubble this quantity is,  
     \begin{equation} \label{tension}
    A_1  \, = \, \frac{1}{6} f^3 \sqrt{\alpha}\, = \, \frac{1}{6} \frac{m^3}{\alpha}.
  \end{equation} 
  The radius of the critical bubble, $r=R$, is found by  
extremizing (\ref{A3}) and it is given by, 
      \begin{equation} \label{RAcrit}
  R \,  = \, \frac{2A_1}{\epsilon} \, = \, \frac{m}{3\mu^2} .
\end{equation} 
   Substituting this into  (\ref{A3}), we get the following value for the energy of the critical bubble, 
      \begin{equation} \label{EA3}
   E_{bub}  \, = \, \frac{16 \pi A_1^3}{3 \epsilon^2} \,  = \, \frac{2 \pi}{81} \frac{m^5}{\mu^4}\frac{1}{\alpha} \,. 
  \end{equation} 
   The action of the corresponding instanton is, 
      \begin{equation} \label{AA3}
   A_{inst}  \, = \, \frac{16 \pi A_1^3}{3 \epsilon^2} \, = \, \frac{2 \pi}{81} \frac{m^5}{\mu^4 T}\frac{1}{\alpha}  
  \end{equation} 

 So far, we gave the standard analysis of the bubble nucleation
along the lines of  \cite{Linde:1981zj}. 
  However, an important novelty in the present 
  case is that both the critical bubble as well as the corresponding 
Euclidean instanton are exponentially degenerate. This is due to the presence of $\sim N$ Goldstone modes of spontaneously broken $SU(N)$ symmetry. 
 This symmetry is spontaneously broken both by the bubble 
as well as by the related instanton. 
 
 Now, the critical bubble becomes a saturon for 
$N \sim 1/\alpha$ and $R \sim 1/m$. Since species 
are in equilibrium, this also implies  
$T \sim m \sim \mu$.    Extrapolating the thin-wall expressions 
(\ref{EA3}) and (\ref{AA3}) to this regime, we see that, simultaneously with the critical bubble,  the entropy bounds 
(\ref{Smax1}) - (\ref{Smax3}) are saturated by the Euclidean instanton.      

At the saturation point the action of the instanton becomes equal to its microstate entropy.
\begin{equation} \label{ASinst} 
  A_{inst} =  S_{inst} \sim \frac{1}{\alpha} 
\end{equation} 
   The instanton entropy provides an additional 
exponential enhancement factor in the transition rate,     \begin{equation} \label{GammaAS}
  \Gamma(T) \sim  {\rm e}^{- A_{inst}  + S_{inst}}\,.
\end{equation}  
 At the saturation point, this factor compensates 
the exponential suppression due to the action 
$A_{inst}$.  At this point the transition probability 
becomes unsuppressed. The result of the transition is 
the creation of the saturon bubble.  
That is,  the Euclidean instanton and the critical bubble 
    that it materializes, become saturated
       simultaneously.                          
    
   Finally, we note that the bubble nucleation process preserves the translation invariance of the thermal bath. This is regardless of saturation of the bubbles and of corresponding instantons. 
  The bubbles are produced in a translation-invariant quantum superposition.  The state is classicalized by inhomogeneities.

   \subsection{Summary and outlook}
   
   The purpose of the present paper is to create awareness about 
 certain potentially-interesting cosmological implications of 
 saturons \cite{Dvali:2020wqi}. 
 These macroscopic objects can come in various forms of 
 bound states such as solitons, baryons
and other creatures \cite{Dvali:2020wqi, Dvali:2019jjw, Dvali:2019ulr,   
Dvali:2021jto, Dvali:2021rlf, Dvali:2021tez, Dvali:2021ofp}.  
  The defining property 
 is that saturons carry a maximal microstate entropy within the validity domain of the
quantum field theoretic description. 
 This feature is quantified by saturation of  the bounds (\ref{Smax1}) -(\ref{Smax3}).

   Due to this property, saturons can be produced in the early Universe via a quantum transition process from the homogeneous thermal bath.      
This production mechanism is not operative for other macroscopic objects, such as black holes or ordinary solitons, albeit due to distinct 
reasons.   

    The thermal production of ordinary solitons is exponentially suppressed.  An example is provided by the suppressed 
bubble nucleation at finite temperature \cite{Linde:1981zj}. 
We showed that this effect can be understood in terms 
of an universal phenomenon of the entropy suppression  
which applies to generic macroscopic objects such as solitons.   
The entropy of an ordinary soliton (e.g., the vacuum bubble
of the previous paper) is typically very low as compared to the entropy of a thermal bath of equal volume and energy.
 It is therefore insufficient for overcoming the exponential 
suppression of the process.

   The story with black holes is different.  The Bekenstein-Hawking 
   entropy of a black hole (\ref{BHe}) \cite{BekE} does saturate the 
   bounds (\ref{Smax1}) - (\ref{Smax3}).   Due to this and other fundamental similarities,  it is clear that black holes represent 
   a special category of saturons \cite{Dvali:2020wqi}.  
    In fact, a black hole is a saturon that is predominantly composed 
   by $N$ gravitons that are held together by gravity 
   \cite{Dvali:2011aa}.    
    
    Despite this similarity, the thermal formation 
    of black holes in a homogeneous radiation dominated Universe  is highly suppressed. This is because, the very same gravitational interaction that keeps black hole in tact,  is at the same time responsible for the cosmological evolution of the Universe.   
   Due to this, a would-be black hole matches the energy  
and size of the thermal radiation only if it covers the entire Hubble patch. However, at this scale the entropy is fully dominated by the Gibbons-Hawking 
entropy (\ref{GH}) \cite{Gibbons:1977mu} 
 of the Hubble patch. Correspondingly, no black hole forms.

 In contrast, the non-gravitational 
 saturons can bypass both difficulties.  On one hand, 
 just like black holes, they carry the maximal entropy. 
  On the other hand, they can match the energy and entropy 
  of the radiation sphere at sub-Horizon scales. 
  Thereby,  saturons  can be created as a result of a quantum transition 
  directly from a homogeneous thermal bath.
  
 Due to an efficient production in the radiation epoch, saturons can serve as viable dark matter candidates.
  We have considered two versions of their cosmological production,
  referred to as ``freeze-out" and ``freeze-in" scenarios. 
  The terminology, which we borrowed from ordinary particle dark matter discussions, refers to the initial level of thermal 
 equilibration of the species that after transition compose a saturon. 
 The rest of physics 
 is fundamentally different, since saturons are collective phenomena  
that never come to thermal equilibrium. 
    We gave qualitative discussions of both scenarios within  
explicit models of saturons.   
 
 The maximal entropy endows saturons with some out of ordinary  
properties.   For example, the superheavy 
macroscopic saturons can be created at very law temperatures. Their interactions and mergers can create the potentially observable signatures such as gravitational waves.

It has been shown \cite{Dvali:2019ulr} that the concept of entropy saturation generalizes to Euclidean configurations such as instantons. 
The entropy bounds (\ref{Smax1})-(\ref{Smax3}) are well-defined 
for such entities.  The instanton entropy emerges from  
a microstate degeneracy of its action.   For the  saturated 
instanton, {\it i}-saturon, the entropy and action are equal.  
The instanton entropy 
provides an additional enhancement factor in 
the transition rate which compensates the suppression due 
to the action. Thereby, the saturation marks the regime in which the 
rate becomes unsuppressed.   Its further increase 
is not possible,  due to invalidation of the 
quantum field theoretic description. 
 
We have shown that  an unsuppressed thermal nucleation 
of a saturon bubble can be understood in terms of a saturated instanton.  
That is, the saturon vacuum bubble is created by a saturated thermal instanton.  In other words, existence of a saturon bubble
is correlated with the existence of an Euclidean 
{\it i}-saturon.  
This is the case even if no stable saturon exists 
at $T=0$.  
 This effect can have important implications for various phase transitions. In particular, it can affect the thermal history 
 of the Universe. 
 It also applies to creation of defect during inflation,  
 discussed by Basu, Guth and Vilenkin \cite{inflation}. 
 The effect of saturation, must enhance the rate of such transitions.

   It is known that  for certain parameter regimes the
 spontaneously broken symmetries are never restored at high temperatures \cite{Weinberg:1974hy, Mohapatra:1979qt, Mohapatra:1979vr}. Instead of vanishing above a certain critical point,  the order parameter grows 
 with increasing temperature.  In such cosmologies, due to the absence of phase transitions,  the topological defects
cannot form via the Kibble mechanism. 
 Hence, 
their quantum production from the thermal bath remains the only 
channel, which for ordinary defects is exponentially suppressed.  
Based on this, in  \cite{Dvali:1995cc, Dvali:1996zr} and \cite{Dvali:1995cj} it 
was argued that symmetry non-restoration eliminates the  
cosmological problems related with domain walls \cite{Zeldovich:1974uw} and monopoles \cite{Mon1,Mon2}.  
 
  As we discussed, saturons avoid the suppression of 
  thermal production, thanks to 
  their entropy.  Due to this, the saturated defects, such as, 
  e.g., magnetic monopoles or the vacuum bubbles, can be 
  produced thermally even in the absence of the Kibble mechanism.
   This can reintroduce the monopole problem in cosmological scenarios with symmetry non-restoration. 
   
     A related interesting question is whether 
  saturated cosmic strings of cosmological importance can be produced via the thermal quantum transition described in this paper. 
  
  It has been argued recently \cite{Dvali:2021ooc} 
that QCD contains
a saturon state  in form of the  Color Glass Condensate \cite{Gelis:2010nm}.
 The present analysis indicates that such a state must have been formed in the early Universe.  Cosmological implications of it remain  to be understood. 
 
Other potentially observable signatures of saturons concern
 the collider experiments. 
  Due to their  maximal microstate entropy, 
saturons with masses up to few TeV
can be produced in high energy 
collisions at particle accelerators such as LHC.  Again, this distinguishes them 
from ordinary solitons with low entropy which are subject to 
an exponential suppression.  
 This distinction also enables for  
direct collider searches for certain categories of saturon dark matter.  
 Some discussions of the generic properties of
 saturon production in particle collisions  can be found in \cite{Dvali:2020wqi}.

 The present paper gives some glimpses of  saturon cosmology.  The discussed examples are 
 indicative and the estimates are qualitative, often relying on
intuitive arguments. Nevertheless, the picture appears to be physically 
sound and interesting from the cosmological perspective. \\

{\textsl{\bf Acknowledgments}}\;---\;
We thank Akaki Rusetsky for useful discussions on aspects 
of large-$N$ and Goran Senjanovi\'c for discussions on symmetry non-restoration. 
This work was supported in part by the Humboldt Foundation under Humboldt Professorship Award, by the Deutsche Forschungsgemeinschaft (DFG, German Research Foundation) under Germany's Excellence Strategy - EXC-2111 - 390814868, and Germany's Excellence Strategy under Excellence Cluster Origins.

\end{document}